\documentclass[letterpaper, 10 pt, conference]{ieeeconf}  

\IEEEoverridecommandlockouts                              

\usepackage{graphics} 
\usepackage{epsfig} 
\usepackage{mathptmx} 
\usepackage{amsmath} 
\usepackage{amssymb}  
\usepackage{hyperref}
\usepackage{svg}
\usepackage{booktabs}
\usepackage{multirow}
\usepackage{graphicx}
\usepackage{subcaption}
\usepackage[font=small]{caption}
\usepackage{tabularx}

\title{\LARGE \bf
Dynamics-Invariant Quadrotor Control using Scale-Aware Deep Reinforcement Learning }

\author{Varad Vaidya$^{1}$ and Jishnu Keshavan$^{2}$
\thanks{The code is open-sourced on \href{https://github.com/varadVaidya/adapt-drones}{GitHub}}
\thanks{*This work is supported by SERB CORE grant.}
\thanks{$^{1}$Varad Vaidya ({\tt{\footnotesize varadmandar@iisc.ac.in}}) is with Robert Bosch Centre for Cyber Physical Systems, Indian Institute of Science, Bangalore, India}%
\thanks{$^{2}$Jishnu Keshavan ({\tt{\footnotesize kjishnu@iisc.ac.in}}) is with the Department of Mechanical Engineering, Indian Institute of Science, Bangalore, India}%
}

\begin{document}
\thispagestyle{empty}
\pagestyle{empty}
\maketitle

\begin{abstract}
Due to dynamic variations such as changing payload, aerodynamic disturbances, and varying platforms, a robust solution for quadrotor trajectory tracking remains challenging. To address these challenges, we present a deep reinforcement learning (DRL) framework that achieves physical dynamics invariance by directly optimizing force/torque inputs, eliminating the need for traditional intermediate control layers. Our architecture integrates a temporal trajectory encoder, which processes finite-horizon reference positions/velocities, with a latent dynamics encoder trained on historical state-action pairs to model platform-specific characteristics. Additionally, we introduce scale-aware dynamics randomization parameterized by the quadrotor's arm length, enabling our approach to maintain stability across drones spanning from 30g to 2.1kg and outperform other DRL baselines by 85\% in tracking accuracy. Extensive real-world validation of our approach  on the Crazyflie 2.1 quadrotor, encompassing over 200 flights, demonstrates robust adaptation to wind, ground effects, and swinging payloads while achieving less than 0.05m RMSE at speeds up to 2.0 m/s. This work introduces a universal quadrotor control paradigm that compensates for dynamic discrepancies across varied conditions and scales, paving the way for more resilient aerial systems.

\end{abstract}
\section{Introduction}
\begin{figure}[t]
    \centering
    \includegraphics{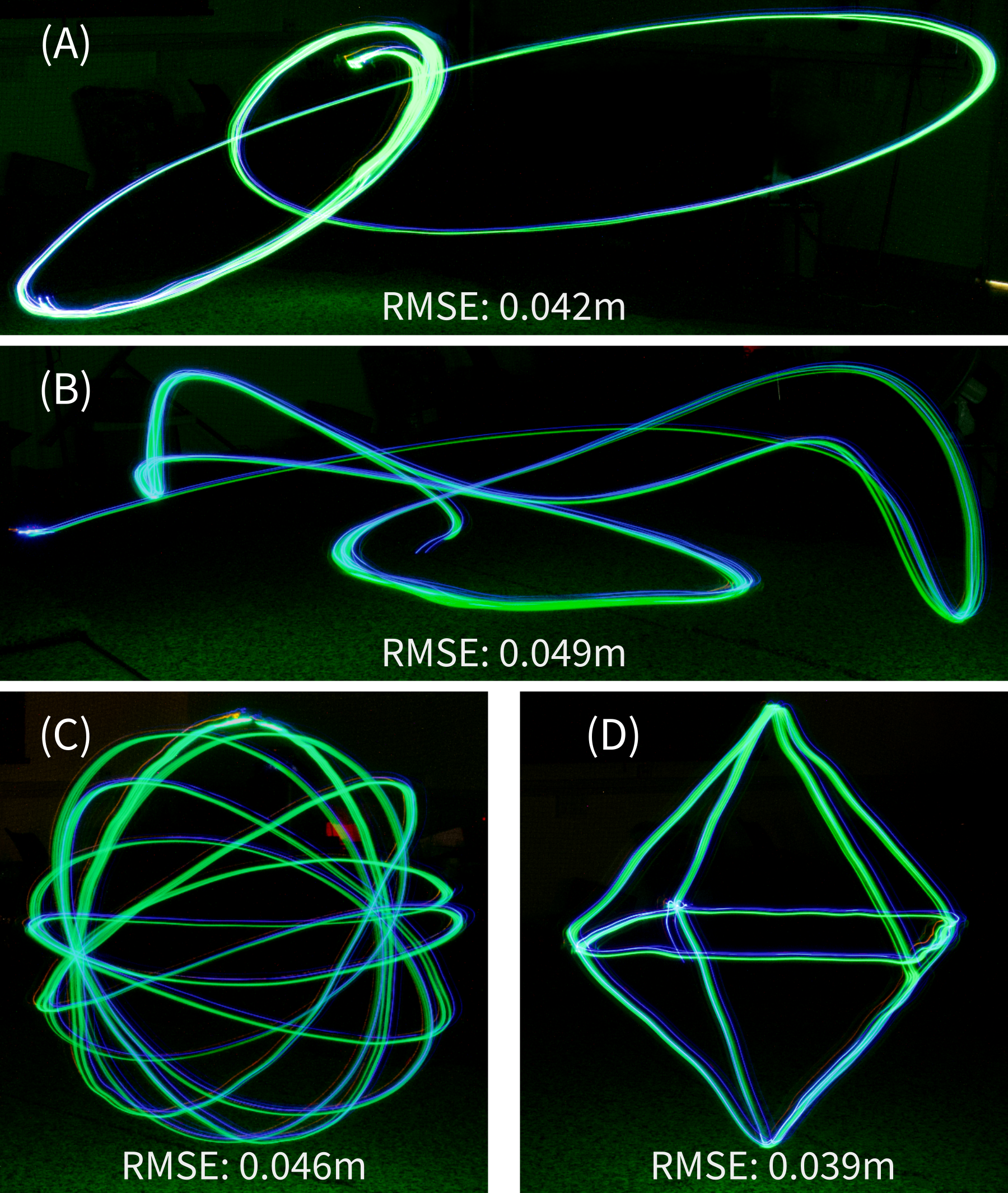}
    \caption{Long exposure images of the Bitcraze Crazyflie 2.1 quadrotor tracking various trajectories in experiments, using the proposed DRL-based dynamics invariant control scheme. Trajectories (A), (B), (C) and (D) are the ``Butterfly", ``Random Spline", ``Satellite Orbit" and ``Octahedron" respectively with the tracking RMSE mentioned below. The baseline trajectories are given in Fig.\,\ref{fig:error_plot}.}
    \label{fig:main_fig}
    \vspace{-17.5pt}
\end{figure}
Quadrotor Unmanned Aerial Vehicles (UAVs) have emerged as versatile platforms for a broad range of applications, from aerial inspection and search-and-rescue operations to precision agriculture and environmental monitoring \cite{rescue}. Despite notable progress in autonomous flight, robust trajectory tracking remains a central challenge due to the intricate interplay of dynamic factors such as payload variations, aerodynamic disturbances, and environmental perturbations.%

Traditional controllers, including cascaded PID and Linear Quadratic Regulators (LQRs), rely on heavy manual tuning or precise system identification and struggle to adapt to unmodeled dynamics. While model predictive control (MPC) improves robustness through receding-horizon optimization, its computational overhead and dependency on accurate dynamics models limit real-time performance in dynamic environments.
In contrast, deep reinforcement learning (DRL) has emerged as a promising alternative by learning control policies directly from interaction data \cite{huttercontrol,championrl}, offering greater adaptability and robustness. However, many DRL approaches encounter a significant sim-to-real gap, with performance deteriorating when deployed under unseen dynamics or across drones with vastly different physical properties often necessitating extensive re-engineering \cite{simplefly}. Additionally, several learning-based methods generate high-level control commands that still depend on quadrotor-specific tuning to convert these outputs into effective motor actuation \cite{kaufmann2022benchmarkcomparisonlearnedcontrol}. Consequently, the development of a universal control framework that seamlessly adapts to diverse scales and environmental disturbances remains an open and pressing research problem.

To address these challenges, we propose a novel deep reinforcement learning (DRL)-based framework that eliminates the need for traditional intermediate control layers by directly optimizing force/torque inputs. Our approach introduces two key innovations: (1) a temporal trajectory encoder that processes finite-horizon reference positions and velocities to guide the quadrotor's motion effectively, and (2) a latent dynamics encoder trained on historical state-action pairs to capture platform-specific characteristics. By integrating these components with scale-aware dynamics randomization parameterized by the quadrotor's arm length, our framework achieves invariance to physical dynamics across a wide range of drone scales—from 30g micro-drones to 2.1kg platforms. This enables the system to maintain stability and high tracking performance despite variations in payloads, aerodynamic disturbances, or other external factors.

\section{Related Work}

This section provides an overview of related work in adaptive quadrotor control to varying dynamics, encompassing both traditional and learning-based methods.
\subsection{Traditional Adaptive Control}

Traditional adaptive control aims to regulate quadrotor dynamics to minimize trajectory tracking errors. Recent works have have explored the use of adaptive \(\mathcal{L}_1\) controllers combined with Model Predictive Control (MPC) \cite{l1mpc} or sampling-based Model Predictive Path Integral (MPPI) control \cite{l1mmpi} to handle model uncertainties effectively. These methods offer rapid adaptation with theoretical guarantees but require explicit reference dynamics parameters, often assuming linear models. Consequently, such controllers lack scale invariance and are designed more for handling small disturbances or unmodeled dynamics rather than adapting across a wide range of quadrotor configurations.

\subsection{Learning-Based Adaptive Control}

Deep reinforcement learning (DRL) has achieved impressive results in quadrotor stabilization \cite{eschmann2024learning} and agile flight \cite{championrl}. Hybrid approaches combining DRL with model-based control \cite{romero2025actorcriticmodelpredictivecontrol} leverage the robustness of model-based techniques while retaining the adaptability of learning-based policies. Imitation learning has also been employed for policy guidance in constrained settings \cite{tubempc, deepdronerace}. Online adaptation modules, such as those in \cite{neuralfly, huang2023dattdeepadaptivetrajectory}, enable policies to respond to disturbances and model mismatches in real time. However, these methods typically remain limited to the quadrotor dynamics encountered during training.

Model-free reinforcement learning has demonstrated promising generalization capabilities. For instance, \cite{molchanov2019simtomultirealtransferlowlevelrobust} successfully transferred a simple hovering policy across multiple drone configurations. Inspired by rapid motor adaptation techniques \cite{kumar2021rmarapidmotoradaptation}, \cite{zhang2023learning} developed a policy capable of controlling quadrotors with vastly different scales through naive domain randomization. However, their approach assumes uniform sampling of dynamic parameters, whereas quadrotor dynamics scale with arm length \cite{Powers2015}. Furthermore, it does not address trajectory or setpoint tracking explicitly.

A parallel effort \cite{zhang2024} introduced dynamics randomization that scales with arm length but still applies uniform scaling of dynamical properties across the considered range. In contrast, our approach incorporates the spread of real-world quadrotor properties, accounting for diverse builds such as FPV drones, long-range drones, and hover-based research platforms.

Another limitation of existing approaches \cite{zhang2023learning, huang2023dattdeepadaptivetrajectory, deepdronerace} is their reliance on Combined Thrust and Body Rate (CTBR) commands. While common in DRL control \cite{kaufmann2022benchmarkcomparisonlearnedcontrol}, CTBR is primarily influenced by mass and nonlinear rotational kinematics \cite{eschmann2024learning}, necessitating quadrotor-specific tuning for effective control. A universal controller should eliminate this requirement by incorporating inertial effects explicitly, using Combined Thrust and Body Torques (CTBT) or Single Rotor Thrusts \cite{eschmann2024learning}.

Thus, in the pursuit of a universal, dynamics-invariant quadrotor controller, we make the following contributions:

\begin{figure}
    \centering
        \includesvg[inkscapelatex=false,width=\columnwidth]{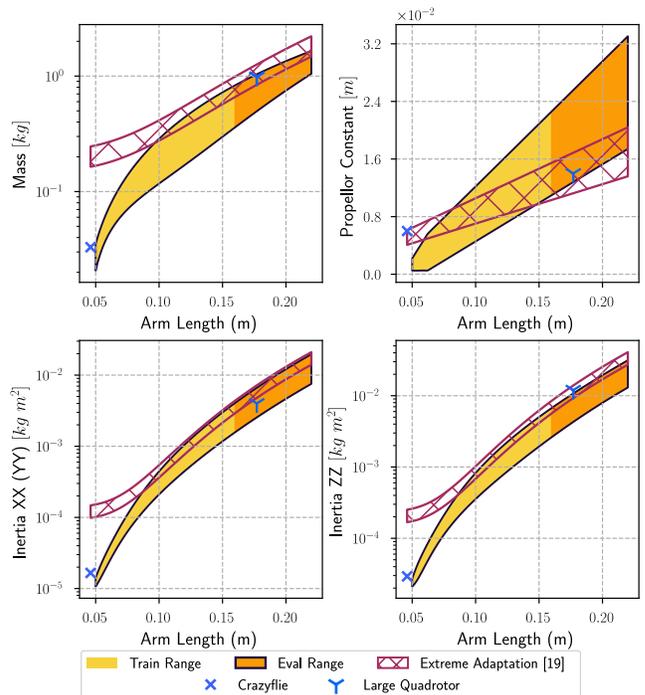}
    \caption{\textbf{Scale-Based Dynamics Randomization}: The training and evaluation domains of our method are shown alongside the parameter space covered by Extreme Adaptation \cite{zhang2024} as functions of arm length. Our approach spans a significantly broader range of dynamic parameters. The Crazyflie 2.1 and ``Large Quadrotor" \cite{zhang2024}, both evaluated in our experiments, lie outside our training range and are indicated for reference.}
    \vspace{-20.5pt}
    \label{fig:scaling}
\end{figure}
\begin{itemize}
\item \textbf{Scale-Based Dynamics Randomization}: We leverage quadrotor dynamic parameters from diverse sources \cite{zhang2023learning, championrl, Powers2015} to capture a wide spectrum of real-world quadrotor designs. Our approach enforces domain randomization based on arm length, ensuring adaptation across varying platform sizes.

\item \textbf{Using CTBT as Policy Outputs}: By directly outputting forces and torques, our policy operates on the fundamental quadrotor dynamical equations, eliminating the need for platform-specific tuning.

\item \textbf{Trajectory Encoder for Planning}: By utilizing only reference positions and velocities over a receding horizon, our approach exploits differentially flat system dynamics
to maximize quadrotor performance. Unlike \cite{huang2023dattdeepadaptivetrajectory}, our method avoids body frame conversions and attitude dependencies, enabling seamless integration with high-level trajectory planners.
\end{itemize}

\section{Dynamics Invariant Controller}

This section details the proposed dynamics invariant control approach. We begin by introducing the notation used followed by, a scale-based dynamics randomization technique designed to capture a wide range of quadrotor configurations. Subsequently, we describe the architecture of our controller, highlighting key modules for trajectory encoding, dynamics estimation, and adaptation. Finally, we will briefly describe the training procedure and reward design used for Proximal Policy Optimization (PPO).

\subsection{Quadrotor Dynamics}
We model the quadrotor as a free-floating rigid body with mass \(m\) and a diagonal inertia tensor \(J\). Its state is described by the position \(p \in \mathbb{R}^3\), the unit quaternion \(q \in \mathbb{S}^3\) representing attitude, the linear velocity \(v \in \mathbb{R}^3\), and the angular velocity \(\omega \in \mathbb{R}^3\). The system dynamics are given by:
\begin{equation}
    \dot{x} =
    \begin{bmatrix}
        \dot{p} \\ \dot{q} \\ \dot{v} \\ \dot{\omega}
    \end{bmatrix} =
    \begin{bmatrix}
        v \\
        \frac{1}{2}\,q \otimes \begin{bmatrix} 0 \\ \omega \end{bmatrix} \\
        -\,g\,\mathbf{e}_3 + \frac{1}{m}\,q \otimes T_B \otimes q^* \\
        J^{-1}\Bigl(\tau - \omega \times (J\,\omega)\Bigr)
    \end{bmatrix}
\end{equation}
where \(\otimes\) denotes quaternion multiplication, \(q^*\) is the conjugate of \(q\), and \(\mathbf{e}_3\) is the unit vector along the vertical axis. In this formulation, \(T_B\) represents the collective thrust in the body frame, and \(\tau\) denotes the body torques are the control inputs to the system which are derived from the motor speeds \(\Omega_i\). Specifically, the thrust and torque produced by the \(i^{\text{th}}\) motor are expressed as \(T_{B_i} = K_F\,\Omega_i^2,\quad \tau_i = K_m\,\Omega_i^2\)
with \(K_F\) and \(K_M\) being the thrust and torque coefficients, respectively.

\subsection{Scale-Based Dynamics Randomization} 
To enhance the robustness and generalization capability of the proposed controller, we employ a scale-based dynamics randomization strategy during the training process.  This approach is motivated by the observation that many inertial properties of quadrotors scale proportionally with their physical size. Specifically, considering the characteristic arm length \(L\) of the quadrotor, we can approximate the scaling of mass \(m\) and inertia \(I\) as \( m \sim L^3\) and \(I \sim L^5\) respectively \cite{Powers2015}.  While the precise relationship between propeller forces/moments and size involves complex aerodynamics, the ratio of torque coefficient to thrust coefficient can be approximated as \(\frac{k_M}{k_F} \sim L\)\, \cite{Powers2015}.

\begin{figure*}[t]
    \centering
    \includegraphics[width=\linewidth]{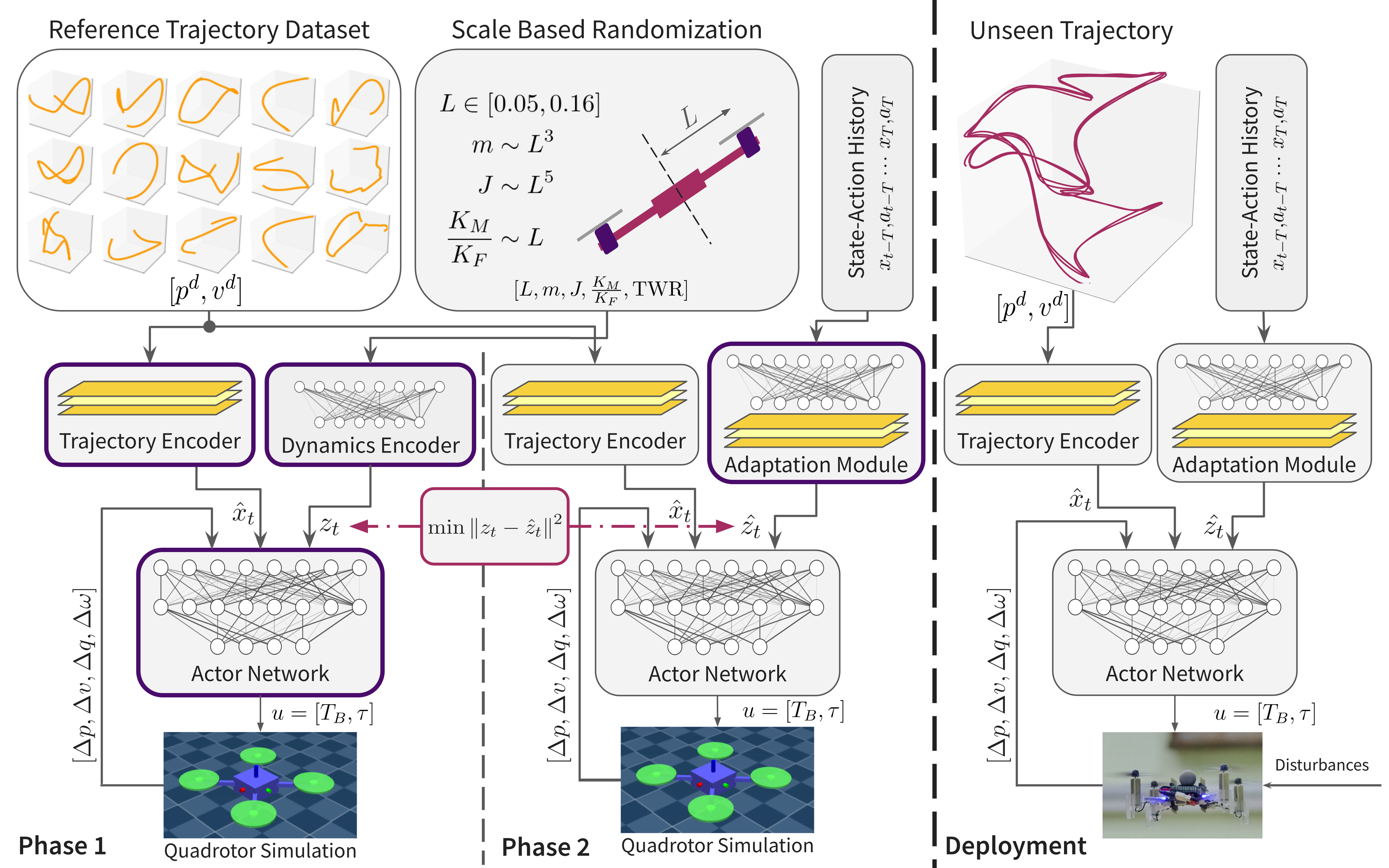}
    \caption{Overview of our two-phase, dynamics-invariant quadrotor controller. In Phase 1, the trajectory encoder, dynamics encoder, and actor network are trained using real-world trajectory data and scale-based randomization. In Phase 2, the adaptation module refines the latent dynamics embedding using supervised learning. Modules trained in each phase are outlined in violet. At deployment, the policy (trajectory encoder, adaptation module, actor network) tracks unseen trajectories without prior knowledge of dynamic parameters. }
    \label{fig:block-diag}
    \vspace{-10pt}
\end{figure*}
These scaling laws allows us to capture a wide variety of quadrotor configurations, as depicted in Fig. \ref{fig:scaling}, while still utilizing low-level commanded thrust and body torque control inputs. Importantly, variations in motor characteristics and overall quadrotor design (FPV and industrial drones, for instance) are addressed by randomizing the thrust-to-weight ratio (TWR) between 2 and 2.75. This range enables the controller to effectively handle a broad class of quadrotors ranging from small aggressive vehicles to large (more docile) platforms.

\subsection{Controller Architecture}
The proposed controller architecture, illustrated in Fig. \ref{fig:block-diag}, builds upon the Rapid Motor Adaptation (RMA) architecture \cite{kumar2021rmarapidmotoradaptation, zhang2023learning} with several key modifications and enhancements.  The core of the controller is a base policy network \(\pi\) that outputs normalized combined thrust and body torques, conditioned on the current pose error.  Critically, the base policy is also conditioned on latent embeddings generated by a trajectory encoder and a dynamics encoder, facilitating adaptation and improved trajectory tracking performance.

\subsubsection{Trajectory Encoder} Drawing inspiration from the receding horizon approach used in Model Predictive Control (MPC), the trajectory encoder module \(h\) encodes deviations from the reference trajectory over a time horizon \(H = 100\) in the inertial reference frame. Specifically, the encoder processes the difference between desired and actual positions and velocities at multiple time steps ahead as:
\begin{equation}
    \hat{x}_t = h( p^d_t - p_t, v^d_t - v_t, \dots, p^d_{t + H} - p_t, v^d_{t+H} - v_t)
\end{equation}
The trajectory encoder's architecture is similar to that described in \cite{huang2023dattdeepadaptivetrajectory}, consisting of three convolutional layers (32 filters, kernel size 3, stride 1) with ReLU activations, followed by a fully connected layer that maps the extracted features to a 32-dimensional embedding.  This embedding captures the anticipated trajectory error, enabling the policy to proactively adjust control actions.

\subsubsection{Dynamics Encoder and Adaptation Module} The dynamics encoder \(\phi\) maps a \(\mathbb{R}^{10}\) vector representing the quadrotor's dynamics properties into an 8-dimensional embedding.  The input vector consists of the quadrotor mass \(m \in \mathbb{R}^1\), inertia \(J\in \mathbb{R}^3\), arm length \(L\in \mathbb{R}^1\), thrust-to-weight ratio \(\text{TWR} \in \mathbb{R}^1 \), propeller constant ratio \(\frac{K_M}{K_F} \in \mathbb{R}^1 \), and wind velocity \(v_w \in \mathbb{R}^3\).  The dynamics encoder comprises two hidden layers with 64 neurons each, using $\tanh(\cdot)$ activation functions to project the input features into the 8-dimensional representation as:
\begin{equation}
    z_t = \phi (m, J, L,\text{TWR},\frac{K_M}{K_F},v_w)
\end{equation}
Following the approach in \cite{kumar2021rmarapidmotoradaptation, zhang2023learning}, we employ an adaptation module to estimate the dynamics encoder embedding \(\hat{z}_t\) from the history of state-action pairs. Specifically, the adaptation network processes the previous 50 timesteps using a three-layer MLP for initial feature extraction, followed by a three-layer 1D CNN with 64 filters per layer, and a final linear layer to output the predicted encoding. The predicted dynamics embedding \(\hat{z}_t\) is then used as a condition for the base policy \(\pi\), allowing the controller to adapt to changing dynamics.
\subsubsection{Base Policy} The base policy network (actor) is a fully connected network with three hidden layers of 64 neurons each, employing \(\tanh(\cdot)\) activations. It takes the pose error, trajectory encoder embedding \(\hat{x}_t\), and predicted dynamics embedding \(\hat{z}_t\) as input and produces a 4-dimensional action representation. These actions represent normalized thrust and torques in the body frame of the quadrotor.

\subsection{Training and Reward Design}

The proposed dynamics-invariant control framework is trained end-to-end in a simulated environment using Proximal Policy Optimization (PPO) \cite{schulman2017proximal}. The training procedure jointly optimizes the base policy, trajectory encoder, and dynamics encoder, enabling the system to learn robust control strategies adaptable to varying dynamic conditions.  The subsequent subsections detail the simulation environment, training process, and reward function design.

\subsubsection{Simulation Environment and Training Procedure}

We utilize the MuJoCo physics engine \cite{todorov2012mujoco} to create a high-fidelity simulation of the quadrotor dynamics.  Training is conducted for 100 million steps across 64 parallel environments to accelerate learning.  Each training episode is set to a duration of 6 seconds, with the simulation running at 100 Hz. The actor and critic networks share a similar architecture, with the critic outputting a single value representing the state value estimate. We warm-start the simulation by uniformly sampling initial positions and velocities from a defined neighborhood around the reference trajectory, while attitude and angular velocities are set arbitrarily. Episodes are terminated if the quadrotor crashes (defined as any part of the quadrotor making contact with the ground) or exceeds predefined boundaries.

A crucial aspect of our training methodology is the use of real-world trajectory data to train the trajectory encoder, rather than relying solely on synthetic trajectories. We leverage the dataset from \cite{pitcn}, resampling eight-second trajectory windows at 100 Hz and adjusting the mean speed uniformly between 1.0 and 1.25 m/s. This allows the trajectory encoder to learn representations that are grounded in realistic flight dynamics.

To enhance simulation fidelity and facilitate sim-to-real transfer, we enable the fluid forces model in MuJoCo, simulating the effects of air drag and viscous resistance. Furthermore, we introduce random wind gusts to the quadrotor, sampled from a distribution \(\mathcal{U}(0,1.5) + \mathcal{N} (0,0.01)\), with wind speeds capped at 2.0 m/s to maintain simulation stability. We also inject noise into the quadrotor's state at each time step to improve robustness: \(p \sim \mathcal{N} (0,0.01), v \sim \mathcal{N} (0,0.01) , q \sim \mathcal{N} (0,0.005), \omega \sim \mathcal{N} (0,0.001) \). Finally, at the beginning of each episode, the quadrotor's physical properties are randomized, where the arm length is sampled which is then used to calculate the mass, inertia, and propeller constant as per Fig.\,\ref{fig:scaling}. 
Since the principle Moments of Inertia, of any rigid body satisfy the triangle inequality (i.e \(J_{xx} + J_{yy} \geq J_{zz}\) and other permutations) \cite{triangle}, we average out sampled diagonal inertia tensor components which violate this constraint.

\subsubsection{Reward Function Design}

The PPO agent is trained to maximize the discounted return, defined by a carefully designed reward function. We adopt the Gaussian reward structure from DeepMind's \texttt{dm\_control} suite \cite{tassa2018deepmindcontrolsuite}, which incorporates a margin parameter to govern the spread of the reward distribution. By adjusting this margin, we can finely tune how rapidly rewards decay away from the target, thereby controlling the sensitivity of the learning signal and promoting effective training. The total reward at each time step is normalized to lie in the range [0, 1] and is composed of the terms mentioned in Table \ref{tab:reward_functions}. The proximity reward with its small margin is designed to maintain a strong gradient signal even when the quadrotor is near the desired position, facilitating precise trajectory following.

\begin{table}
\centering
\caption{Reward function components with corresponding margins and weights.}
\label{tab:reward_functions}
\begin{tabular}{l l l l}
\hline
\textbf{Reward Term}         & \textbf{Definition}                          & \textbf{Margin} & \textbf{Weight} \\ \hline
Position Error               & \(r_{pos} = \|x^d_t - x_t\|\)                 & 0.75            & 0.50          \\
Proximity Reward             & \(r_{close\_pos} = \|x^d_t - x_t\|\)          & 0.05            & 0.1625        \\
Velocity Error               & \(r_{vel} = \|v^d_t - v_t\|\)                 & 0.50            & 0.20          \\
Angular Velocity Error       & \(r_{\omega} = \|\omega^d_t - \omega_t\|\)     & 0.50            & 0.0625        \\
Action Smoothness            & \(r_{smooth} = \|a_t - a_{t-1}\|\)             & 0.50            & 0.0375        \\
Yaw Error                    & \(r_{yaw} = \|\psi^d_t - \psi_t\|\)            & 0.50            & 0.0375        \\
Crash Penalty                & \(r_{crash} = -100\)                          & --              & --            \\ \hline
\end{tabular}
\vspace{-5pt}
\end{table}
\begin{figure}
    \centering
    \includesvg[inkscapelatex=false,width=\linewidth]{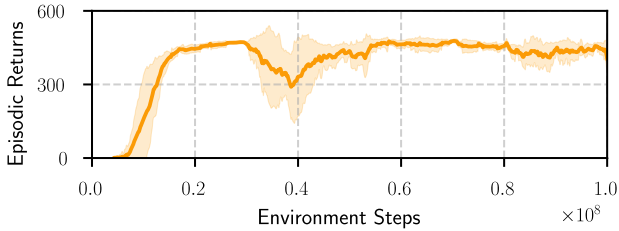}
    \caption{Episodic returns over training iterations, averaged over 5 independent PPO runs with different random seeds, demonstrating the stable convergence of our learning algorithm.}
    \label{fig:returns}
    \vspace{-15pt}
\end{figure}
\begin{figure*}
    \centering
    \includesvg[inkscapelatex=false,width=\linewidth]{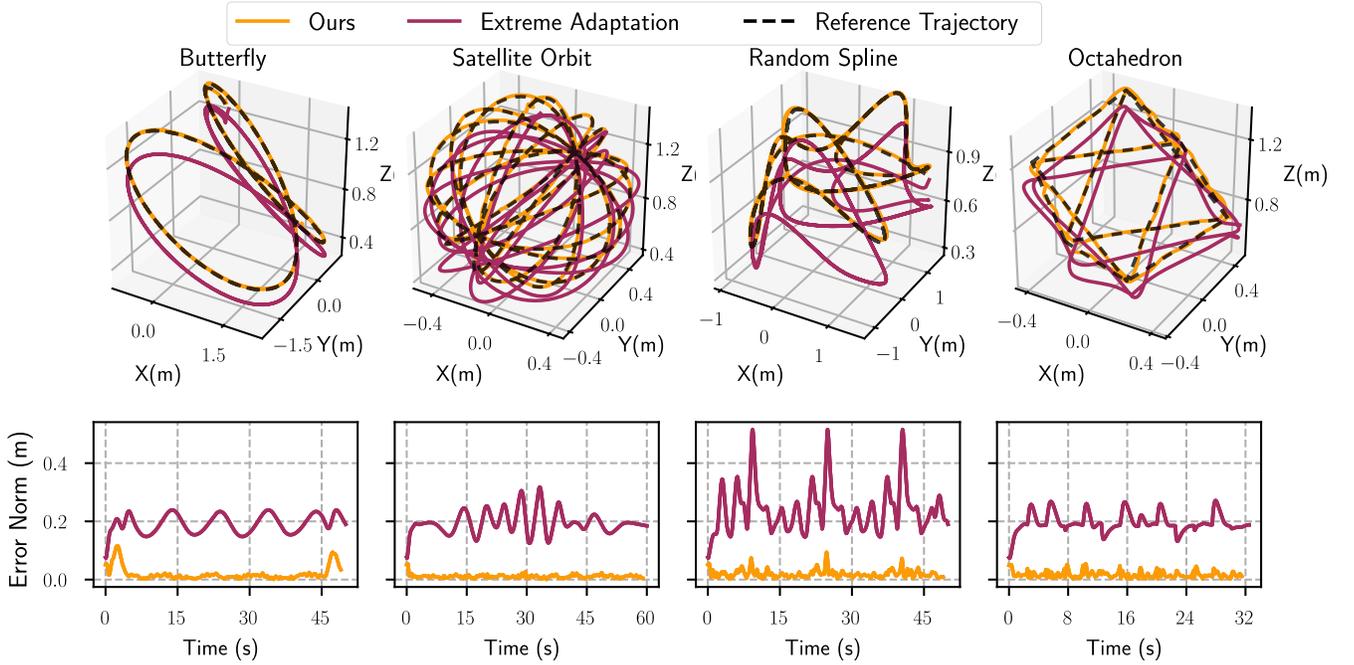}
    \caption{Comparison against the Extreme Adaptation architecture \cite{zhang2024} for trajectory tracking on four complex paths: Butterfly, Satellite Orbit, Random Spline, and Octahedron, using ``Large Quadrotor" (ref. Fig. \ref{fig:scaling}). The top row shows the 3D flight paths generated by the proposed scheme and study in \cite{zhang2024} in relation to the reference path. The bottom row presents the normed position error over time, illustrating our method’s consistently lower tracking error.}
    \label{fig:error_plot}
\end{figure*}

Training of the entire pipeline is undertaken with an Intel i7-12700 4.9GHz workstation equipped with an NVIDIA RTX A4000 graphics card. Episodic returns averaged over 5 runs with different random seeds is shown in Fig. \ref{fig:returns}.
\section{Simulation Results}
\subsection{Benchmark Comparison}
We begin by evaluating our controller under our custom quadrotor dynamics scaling. For benchmarking purposes, we compare our results against a ``Nominal MPC" strategy, which employs a quadratic optimization problem solved via Sequential Quadratic Programming (SQP) in a Real-Time Iteration (RTI) scheme, as described in \cite{torrente2021data}. Notably, the Nominal MPC employs a non-adaptive nominal dynamics model for planning, thereby representing a traditional model-based controller with erroneous system dynamics. Additionally, we include a ``Ground Truth MPC” that operates under ideal conditions (i.e., no noise or wind disturbances) to serve as an optimal performance reference. To ensure a fair comparison, both MPC controllers utilize a prediction horizon equivalent to the trajectory window employed by the trajectory encoder in our approach, which is chosen as 1 second in this study.

Four distinct trajectories are used to evaluate controller performance.  The ``Butterfly" trajectory closely resembles the training distribution, while ``Satellite Orbit" and ``Random Spline" challenge the controller with high-speed maneuvers reaching up to 2.1 m/s. The ``Octahedron" further tests the handling of non-smooth trajectory.  Table \ref{tab:sim_no_extreme} presents the comparative results, demonstrating the superior performance of our proposed controller against the ``Nominal MPC" across all tested arm lengths (and thus dynamics) and trajectory types. This improved performance is due to the trajectory encoder's ability to effectively capture state errors over a receding horizon, facilitating accurate trajectory tracking.  This, combined with the adaptation module's inference of the correct dynamics for appropriate control actions, enables robust performance across diverse conditions.

\subsection{Robustness Analysis: Case Studies} To evaluate the robustness of our proposed controller, we conduct three case studies. The first case assesses the controller’s adaptability to different dynamics scaling laws, demonstrating effective control of quadrotor configurations not included during policy training. The second  examines the adaptation module's ability to handle wind disturbances. Finally, the third study evaluates the trajectory encoder's generalization by testing the maximum achievable speed. A fixed thrust-to-weight ratio of 2.5 is used in both the wind disturbance and maximum speed experiments.
\begin{table*}[t]
\centering
\caption{Evaluation of trajectory tracking over two different dynamics scaling laws for four distinct trajectories. Results are reported in centimeters (cm) and averaged over 10 random seeds. The proposed controller and Extreme Adaptation \cite{zhang2024} were evaluated in a zero-shot manner, without retraining for the changed dynamics. Ground Truth MPC \cite{torrente2021data} acts as an idealized control baseline for the considered receding horizon. Our approach consistently outperforms the baselines with low variance over test runs.}
\label{tab:sim_combined}
\footnotesize
\begin{subtable}[t]{0.48\textwidth}
  \centering
  \caption{\small Trajectory tracking performance using proposed dynamics scaling. Results highlight significantly lower RMSE compared to all other baselines.}
  \begin{tabular}{l c c c c}
    \toprule
    \textbf{\shortstack{Arm\\Length (m)}} & \textbf{\shortstack{Nominal\\MPC}} & \textbf{\shortstack{Extreme\\Adaptation}} & \textbf{\shortstack{Ours \\ \phantom{MPC}}} & \textbf{\shortstack{Ground\\Truth MPC}} \\
    \midrule
    \multicolumn{5}{c}{\textbf{Butterfly}} \\
    \midrule
    0.050 & 15.6$\pm$2.0 &35.8$\pm$0.0& \textbf{2.0$\pm$0.1} & 0.3$\pm$0.0 \\
    0.110 & 7.8$\pm$1.7  &35.8$\pm$0.0& \textbf{2.0$\pm$0.1} & 0.3$\pm$0.0 \\
    0.160 & 4.7$\pm$0.9  &34.8$\pm$4.6& \textbf{2.0$\pm$0.1} & 0.3$\pm$0.0 \\
    0.210 & 3.5$\pm$0.5  &20.0$\pm2.3$& \textbf{2.0$\pm$0.1} & 0.3$\pm$0.0 \\
    \midrule
    \multicolumn{5}{c}{\textbf{Satellite Orbit}} \\
    \midrule
    0.050 & 15.7$\pm$2.9 &41.4$\pm$0.8& \textbf{2.0$\pm$0.1} & 0.4$\pm$0.0 \\
    0.110 & 7.8$\pm$2.0  &41.2$\pm$1.2& \textbf{1.9$\pm$0.1} & 0.4$\pm$0.0 \\
    0.160 & 4.6$\pm$1.0  &39.5$\pm$5.7& \textbf{1.9$\pm$0.1} & 0.4$\pm$0.0 \\
    0.210 & 3.5$\pm$0.5  &24.5$\pm$3.5& \textbf{2.0$\pm$0.1} & 0.4$\pm$0.0 \\
    \midrule
    \multicolumn{5}{c}{\textbf{Random Spline}} \\
    \midrule
    0.050 & 15.1$\pm$2.5 &37.6$\pm$0.3& \textbf{1.2$\pm$0.1} & 0.2$\pm$0.0 \\
    0.110 & 7.6$\pm$2.0  &37.4$\pm$0.3& \textbf{1.2$\pm$0.1} & 0.2$\pm$0.0 \\
    0.160 & 4.6$\pm$1.1  &35.5$\pm$4.7& \textbf{1.2$\pm$0.1} & 0.2$\pm$0.0 \\
    0.210 & 3.4$\pm$0.6  &20.3$\pm$2.4& \textbf{1.3$\pm$0.1} & 0.2$\pm$0.0 \\
        \midrule
    \multicolumn{5}{c}{\textbf{Octahedron}} \\
    \midrule
    0.050 & 16.0$\pm$2.8 &43.4$\pm$0.2& \textbf{1.6$\pm$0.1} & 0.5$\pm$0.0 \\
    0.110 & 7.9$\pm$1.9  &45.4$\pm$0.2& \textbf{1.6$\pm$0.1} & 0.5$\pm$0.0 \\
    0.160 & 4.6$\pm$0.9  &43.7$\pm$5.8& \textbf{1.7$\pm$0.1} & 0.5$\pm$0.0 \\
    0.210 & 3.4$\pm$0.5  &20.2$\pm$2.9& \textbf{1.8$\pm$0.1}& 0.5$\pm$0.0 \\
    
    \bottomrule
  \end{tabular}
  \label{tab:sim_no_extreme}
\end{subtable}
\begin{subtable}[t]{0.48\textwidth}
  \centering
  \caption{\small Trajectory tracking performance under the Extreme Adaptation scaling law \cite{zhang2024} demonstrates that our approach adapts to diverse quadrotor parameters without compromising performance.}
  \begin{tabular}{l c c c c}
    \toprule
    \textbf{\shortstack{Arm\\Length (m)}} & \textbf{\shortstack{Nominal\\MPC}} & \textbf{\shortstack{Extreme\\Adaptation}} & \textbf{\shortstack{Ours \\ \phantom{MPC}}} & \textbf{\shortstack{Ground\\Truth MPC}} \\
    \midrule
    \multicolumn{5}{c}{\textbf{Butterfly}} \\
    \midrule
    0.050 & 10.8$\pm$1.1 & 14.8$\pm$4.1 & \textbf{1.6$\pm$0.0} & 0.3$\pm$0.0 \\
    0.110 & 10.1$\pm$0.8  & 30.9$\pm$7.9 & \textbf{1.7$\pm$0.0} & 0.3$\pm$0.0 \\
    0.160 & 4.9$\pm$0.4  & 29.5$\pm$9.0 & \textbf{1.8$\pm$0.0} & 0.3$\pm$0.0 \\
    0.210 & 3.5$\pm$0.3  & 31.6$\pm$16.4 & \textbf{1.8$\pm$0.0} & 0.3$\pm$0.0 \\
    \midrule
    \multicolumn{5}{c}{\textbf{Satellite Orbit}} \\
    \midrule
    0.050 & 11.1$\pm$1.0 & 17.4$\pm$5.6 & \textbf{1.9$\pm$0.1} & 0.4$\pm$0.0 \\
    0.110 & 9.8$\pm$0.7  & 34.9$\pm$8.6 & \textbf{1.8$\pm$0.1} & 0.4$\pm$0.0 \\
    0.160 & 4.9$\pm$0.4  & 34.9$\pm$10.9 & \textbf{1.8$\pm$0.1} & 0.4$\pm$0.0 \\
    0.210 & 3.4$\pm$0.3  & 41.6$\pm$23.4 & \textbf{3.4$\pm$0.1} & 0.4$\pm$0.0 \\
    \midrule
    \multicolumn{5}{c}{\textbf{Random Spline}} \\
    \midrule
    0.050 & 10.5$\pm$0.8 & 14.9$\pm$4.2 & \textbf{1.1$\pm$0.1} & 0.2$\pm$0.0 \\
    0.110 & 9.9$\pm$0.5  & 32.1$\pm$8.4 & \textbf{1.1$\pm$0.1}& 0.2$\pm$0.0 \\
    0.160 & 4.6$\pm$0.2  & 30.7$\pm$9.6 & \textbf{1.1$\pm$0.1} & 0.2$\pm$0.0 \\
    0.210 & 3.2$\pm$0.2  & 32.9$\pm$18.1 & \textbf{1.1$\pm$0.1} & 0.2$\pm$0.0 \\
        \midrule
    \multicolumn{5}{c}{\textbf{Octahedron}} \\
    \midrule
    0.050 & 10.9$\pm$0.7 & 14.5$\pm$4.3 & \textbf{1.3$\pm$0.1} & 0.5$\pm$0.0 \\
    0.110 & 10.0$\pm$0.6  & 14.5$\pm$4.3 & \textbf{1.4$\pm$0.1} & 0.5$\pm$0.0 \\
    0.160 & 4.8$\pm$0.3  & 35.7$\pm$14.1 & \textbf{1.6$\pm$0.1} & 0.5$\pm$0.0 \\
    0.210 & 3.5$\pm$0.3  & 31.1$\pm$14.7 & \textbf{2.4$\pm$0.1} & 0.5$\pm$0.0 \\
    \bottomrule
  \end{tabular}
  \label{tab:sim_extreme}
\end{subtable}
\end{table*}
\subsubsection{Robustness to Quadrotor Dynamics}
Parallel to our work, \cite{zhang2024} introduced a universal controller based on scale-driven domain randomization, which we refer to here as ``Extreme Adaptation”. While \cite{zhang2024} explored a different range of dynamics, as illustrated in Fig. \ref{fig:scaling}, we evaluated our controller against their proposed dynamics variations without any retraining. Notably, the difference in dynamics parameters, especially at small arm lengths, is substantial as shown in Fig \ref{fig:scaling}. Using the same trajectories and evaluation parameters described earlier, the results in Table \ref{tab:sim_extreme} show that our approach consistently achieves lower tracking error across all lengths and new quadrotor parameters, with up to a 95\% reduction in RMSE.
To further illustrate this performance, Fig \ref{fig:error_plot} showcases the trajectory tracking and corresponding error norms over time for the ``Large Quadrotor" configuration from \cite{zhang2024}, which lies outside our policy's training range. As the figure indicates, our controller maintains accurate trajectory tracking, demonstrating its strong generalization capabilities.
\begin{figure}[t]
    \centering
    \includesvg[inkscapelatex=false,width=\linewidth]{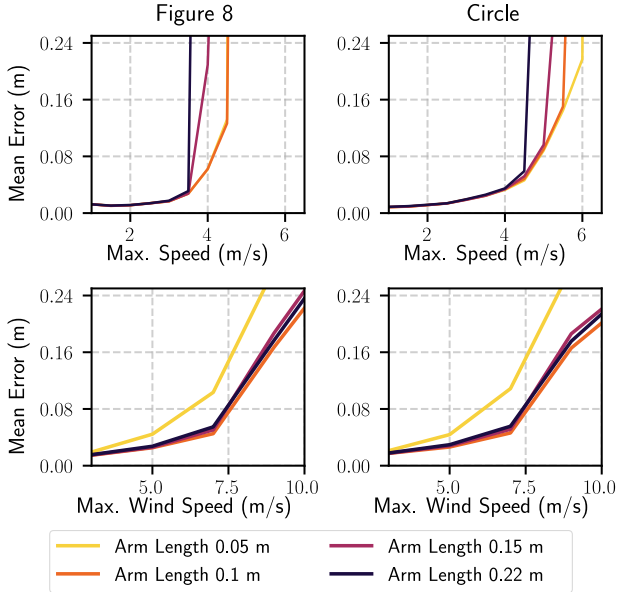}
    \caption{Generalization capabilities of the proposed approach across varying arm lengths. The top row shows the mean tracking error for increasing maximum vehicle speed for both circle and lemniscate (radius 5m) trajectories, while the bottom row illustrates the maximum wind speed tolerated by the quadrotor.}
    \label{fig:max_wind_plot}
    \vspace{-15pt}
\end{figure}

\subsubsection{Resilience to Wind Disturbances}
We evaluate trajectory tracking on circular and Figure-8 trajectories (each with a 5\,m radius) at a nominal speed of 2\,m/s under wind disturbances. This experiment tests the adaptation module’s ability to generalize to wind conditions beyond those encountered during training. As shown in Fig.~\ref{fig:max_wind_plot}, our controller successfully handles wind speeds of up to 10\,m/s, five times the maximum wind speed seen during training.
\subsubsection{Maximum Achievable Speed}
We assess the receding horizon generalization of the trajectory encoder by determining the maximum speed attainable on the same circular and Figure-8 trajectories. These tests are particularly challenging as they require the quadrotor to execute aggressive roll and pitch maneuvers not observed during training. As depicted in Fig.~\ref{fig:max_wind_plot}, our controller achieves speeds of up to 4.5\,m/s, which is three times higher than the speeds present in the training dataset.

\section{Hardware Results}
To validate our controller in a real-world setting, we deployed it on a Crazyflie 2.1 platform equipped with high-thrust motors for enhanced handling. The quadrotor, weighing 39 g, has a thrust-to-weight ratio of 1.9, which lies outside the training range of our controller, as shown in Fig. \ref{fig:scaling}.  For these experiments, we utilized Crazyswarm2 alongside a Qualisys Motion Capture System to track position and velocity data, while attitude information was logged directly from the Crazyflie 2.1.
The hardware evaluation was conducted across several scenarios to assess both baseline performance and robustness. Baseline tests involved tracking a variety of trajectories under nominal conditions, including smooth and non-smooth trajectories. Notably, the inclusion of non-smooth trajectories absent during training, demonstrates the controller’s ability to generalize beyond its training distribution (Fig. \ref{fig:main_fig}).

\begin{figure}
    \centering
    \includegraphics[width=\linewidth]{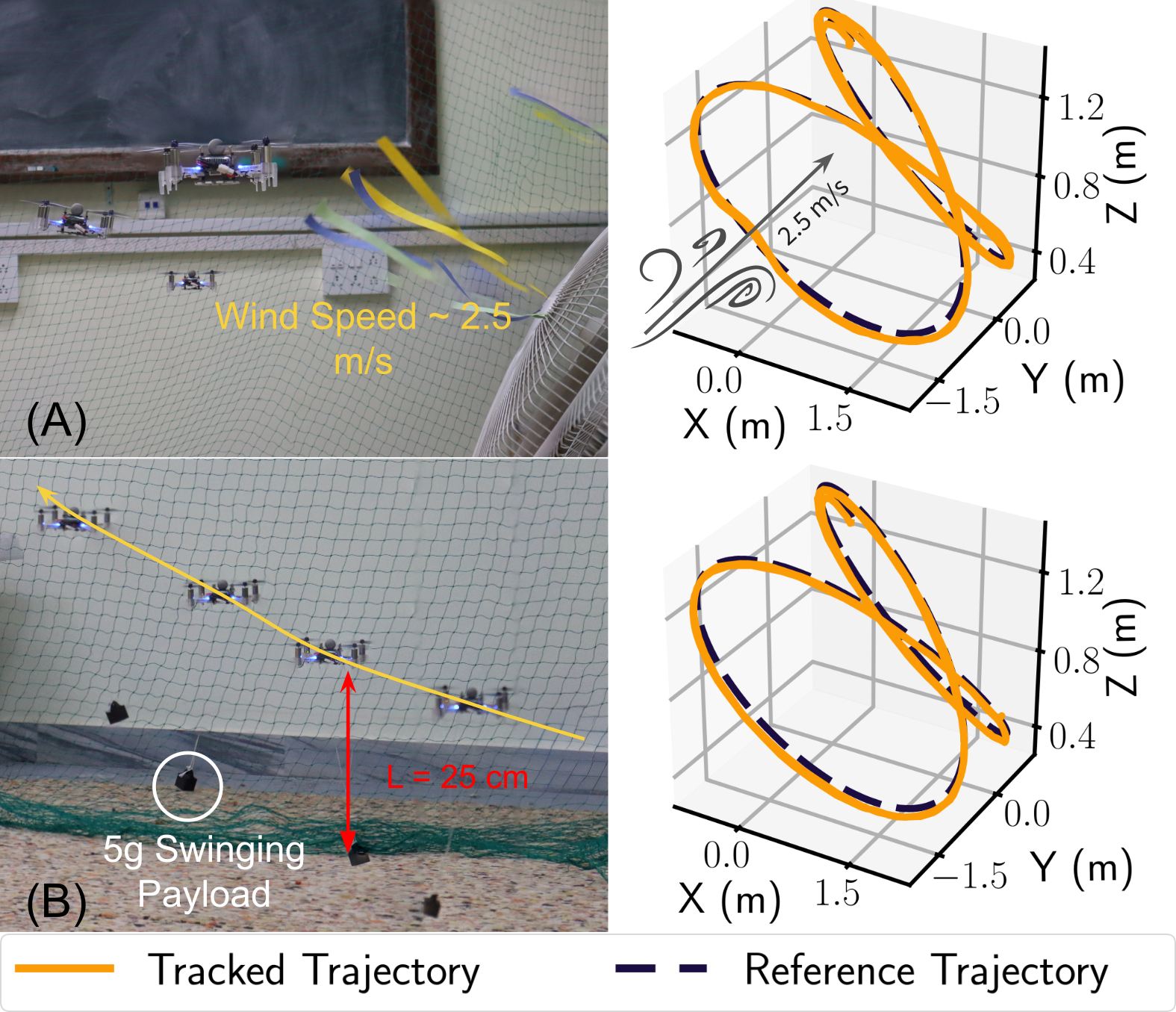}
    \caption{Testing of the proposed controller under (A) wind gust reaching upto 2.5\,m/s tracking ``Butterfly" trajectory. (B) 5\,g swinging payload representing 10\% mass change, tracking ``Butterfly" trajectory. }
    \label{fig:wind_stuff}
    \vspace{-5pt}
\end{figure}

To evaluate robustness further, we subjected the controller to a dynamic wind field with gusts reaching speeds of 2.5 m/s. Additionally, we tested its performance under varying payload conditions by attaching a swinging payload of 5g (representing a 10\% mass increase) suspended from a 25 cm cable (about $5\times$arm length) as shown in  Fig \ref{fig:wind_stuff}. 
Table \ref{tab:trajectory_tracking} summarizes our controller’s performance across various trajectories. Our approach consistently achieves low RMSE values with minimal standard deviation, underscoring both high accuracy and repeatability. Notably, performance on the non-smooth trajectory is comparable to that on smoother paths, demonstrating strong generalization. Under wind disturbances, the average RMSE increased by only 0.016 m, while payload perturbations led to a marginal increase of 0.011m, even as the quadrotor reached speeds of up to 2.1 m/s on the “Random Spline” trajectory, approaching the limits of the system dynamics.
\begin{table}[t]
    \footnotesize 
    \setlength{\tabcolsep}{4pt} 
    \centering
    \caption{Experimental evaluation of trajectory tracking performance under various situations (wind speeds upto 2.5 m/s, 5\,g swinging payload) on Crazyflie 2.1. The reported tracking error metrics are in meters with each metric averaged over 10 runs.}
    \label{tab:trajectory_tracking}
    \begin{tabular*}{\columnwidth}{@{\extracolsep{\fill}}lccc}
        \toprule
        \textbf{Trajectory} & \textbf{Baseline} & \textbf{Wind} & \textbf{Payload} \\
        \midrule
        Butterfly            & $0.044 \pm 0.003$ & $0.062 \pm 0.002$ & $0.048 \pm 0.004$ \\
        Satellite Orbit  &  \( 0.0460 \pm 0.002\) & \(0.0483 \pm 0.004\) & \(0.0490 \pm 0.003\) \\
        Random Spline Loop   & $0.051 \pm 0.002$ & $0.076 \pm 0.003$ & $0.084 \pm 0.003$ \\
        Octahedron        & \(0.0393 \pm 0.002\) & \(0.0432 \pm 0.002\) & \(0.0473 \pm 0.004\) \\
        \bottomrule
    \end{tabular*}
\end{table}
\begin{table}[!t]
    \footnotesize 
    \centering
    \caption{Experimental evaluation of ground effect on tracking the lemniscate trajectory in Fig. \ref{fig:max_wind_plot_1} at different altitudes above ground for different maximum vehicle velocities. The reported tracking error metrics are in meters with each metric averaged over 10 runs.}
    \label{tab:ground_effect}
    \begin{tabular*}{\columnwidth}{@{\extracolsep{\fill}}lccc}
        \toprule
        \textbf{Reference height (m)} & \textbf{0.75 m/s} & \textbf{1 m/s} & \textbf{1.25 m/s} \\
        \midrule
        0.08 & $0.019 \pm 0.002$ & $0.027 \pm 0.001$ & $0.038 \pm 0.001$ \\
        0.10  & $0.026 \pm 0.002$ & $0.035 \pm 0.002$ & $0.045 \pm 0.001$ \\
        $\geq$ 0.20  & $0.035 \pm 0.002$ & $0.042 \pm 0.002$ & $0.054 \pm 0.002$ \\
        \bottomrule
    \end{tabular*}
    \vspace{-10pt}
\end{table}
\begin{figure}[t]
    \centering
    \includesvg[inkscapelatex=false,width=\linewidth]{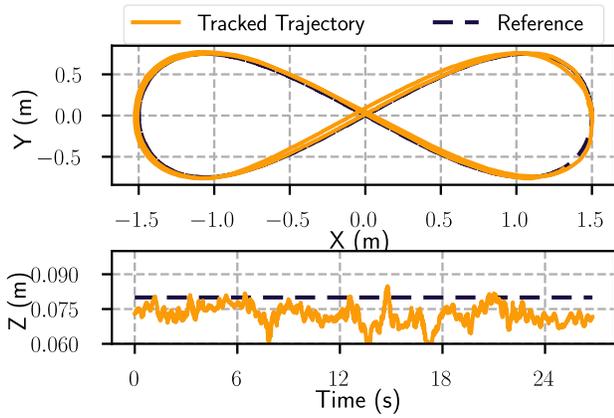}
    \caption{Sample trajectory tracking performance of the Crazyflie 2.1 under ground effect with a maximum speed of 1.25\,m/s. }
    \label{fig:max_wind_plot_1}
    \vspace{-15pt}
\end{figure}
Finally, to assess its response to unmodeled aerodynamic forces, we tracked a lemniscate trajectory in the presence of ground effect at varying speeds and heights \cite{geeffect}. For the same power consumed, ground effect manifests as an increase in thrust generated by the rotors (resulting in improved energy efficiency) close to ground, which renders the problem of achieving accurate tracking at different speeds very challenging. Table \ref{tab:ground_effect} and Fig. \ref{fig:max_wind_plot_1} present tracking results under ground effect at various speeds. It is apparent that tracking performance improves with energy efficiency as the quadrotor gets closer to ground. These findings confirm that the adaptation module effectively compensates for unmodeled aerodynamic forces, as indicated by the minimal degradation in tracking accuracy. Collectively, these experimental results validate that our controller reliably maintains accurate trajectory tracking across a wide range of operating conditions and disturbances.

\section{Conclusion}
This work presents the design of a dynamics-invariant controller capable of accurate trajectory tracking across a wide range of quadrotor dynamics, including unmodeled aerodynamic effects and external disturbances.  Through extensive simulation and hardware experiments, we show that our approach consistently outperforms nominal MPC baselines, and other learning-based universal controllers achieving lower RMSE values across diverse trajectory types, dynamic variations, and real-world conditions. By achieving accurate and resilient control without prior knowledge of specific vehicle parameters, our method paves the way for more versatile and adaptive autonomous aerial systems. In the future, we aim to extend our framework to support more complex aerial vehicles and include vision-based state estimation for better performance in practical applications. Additionally, we also plan to combine model-based control elements with the learned policy to improve safety and interpretability, paving the way for more reliable flight operations.










\bibliographystyle{IEEEtran}
\bibliography{references}

\end{document}